\documentclass[aps,twocolumn,prd,preprintnumbers, superscriptaddress,floatfix, showpacs]{revtex4-1} %if we want to have it in the same format as earlier just insert nofootinbib, in the document class
\usepackage[utf8]{inputenc}
\usepackage{amsmath}
\usepackage{bm}
\usepackage{amsfonts}
\usepackage{graphicx}
\usepackage{epstopdf}
\usepackage{hyperref}
\usepackage{array}
\usepackage{soul}
\usepackage{xcolor}
\usepackage{comment}
\enlargethispage{\baselineskip}

\hypersetup{
     colorlinks   = true,
     citecolor    = blue
}

\newcommand*{\lh}[1]{\lambda_{h}}
 
\newcommand*{\lc}[1]{\lambda_{c}}
\newcommand*{\lo}[1]{\lambda_{0}}

\newcommand{\de}{\mathrm{d}}
\newcommand{\GeV}{\,\mathrm{GeV}}
\newcommand{\MeV}{\,\mathrm{MeV}}

\begin{document}
\title{\bf Echo of the Dark:\\ Gravitational waves from dark SU(3) Yang-Mills theory}
\author{Enrico Morgante}\email[Electronic address: ]{emorgant@uni-mainz.de}
\affiliation{PRISMA+ Cluster of Excellence \& Mainz Institute for Theoretical Physics, Johannes Gutenberg-Universit{\"a}t Mainz, 55099 Mainz, Germany}
\author{Nicklas Ramberg}\email[Electronic address: ]{nramberg@uni-mainz.de}
\affiliation{PRISMA+ Cluster of Excellence \& Mainz Institute for Theoretical Physics, Johannes Gutenberg-Universit{\"a}t Mainz, 55099 Mainz, Germany}
\author{Pedro Schwaller}\email[Electronic address: ]{pedro.schwaller@uni-mainz.de}
\affiliation{PRISMA+ Cluster of Excellence \& Mainz Institute for Theoretical Physics, Johannes Gutenberg-Universit{\"a}t Mainz, 55099 Mainz, Germany}
\date{\today}

\begin{abstract}
We analyze the phase transition in improved holographic QCD to obtain an estimate of the gravitational wave signal emitted in the confinement transition of a pure SU(3) Yang-Mills dark sector. We derive the effective action from holography and show that the energy budget and duration of the phase transition can be calculated with minor errors. These are used as input to obtain a prediction of the gravitational wave signal. To our knowledge, this is the first computation of the gravitational wave signal in a holographic model designated to match lattice data on the thermal properties of pure Yang-Mills. 
\end{abstract}
\begin{flushright}
  MITP-22-088
\end{flushright}
\maketitle

\section{Introduction}
%\vspace*{-1\baselineskip}
%\textbf{Introduction:}
First order phase transitions (FOPT) in the early universe emitting gravitational waves (GWs) are detectable in upcoming experiments \cite{LISA:2017pwj,Janssen:2014dka,Yagi:2011wg}, and would be clear hints for new physics since the electroweak and QCD PTs in the standard model are cross-overs. 
These GWs might allow us to probe the dynamics of otherwise inaccessible dark or hidden sectors~\cite{Schwaller:2015tja,Breitbach:2018ddu,Fairbairn:2019xog}. 
SU(N) Yang-Mills theories are known to feature a color confinement FOPT~\cite{Svetitsky:1982gs,Panero:2009tv}, and they appear in many extensions of the standard model~\cite{Gross:1984dd,Dixon:1985jw,Dixon:1986jc,Acharya:1998pm,Halverson:2015vta,Asadi:2021pwo,Asadi:2021yml,Bai:2013xga,Schwaller:2015gea}. 
These scenarios are minimal in the sense that the confinement scale is their only free parameter, and thus ideal as benchmark models. 
However, due to the strong coupling, nonperturbative methods are needed for quantitative studies of the dynamics.

Here we employ the AdS/CFT correspondence~\cite{Maldacena:1997re,Witten:1998zw} in a bottom-up framework. We use the Improved Holographic QCD model~\cite{Gursoy:2007cb,Gursoy:2007er}, which successfully reproduces lattice data of SU(3) thermodynamics~\cite{Gursoy:2008za,Gursoy:2009jd}, to calculate the equilibrium and quasiequilibrium quantities relevant for GWs.
Previous attempts to study the GW signal in such models include Refs.~\cite{Halverson:2020xpg, Bigazzi:2020phm, Bigazzi:2020avc, Huang:2020mso, Wang:2020zlf, Kang:2021epo, Yamada:2022aax, Yamada:2022imq}.

The outline of this work is as follows. We will start by reviewing Improved Holographic QCD and compute the equilibrium thermodynamics of the model. We will then construct an effective action by using the free energy landscape approach \cite{Creminelli:2001th,vonHarling:2017yew,Baratella:2018pxi,Bigazzi:2020avc,Bigazzi:2020phm}. For the kinetic term of the effective action, we follow the approach of \cite{Baratella:2018pxi} regarding its normalization. We use our effective action to calculate the GW signal by using the LISA Cosmology Working Group \cite{Caprini:2019egz} template for the PT parameters $\beta, \alpha, v_{w}, \kappa(\alpha)$. Finally, we will discuss our results and some future prospects.

\section{Review of Improved Holographic QCD}
% \vspace*{-1\baselineskip}
%\label{sec:IHQCD}
%\smallskip
%\textbf{Review of Improved Holographic QCD:}
Improved Holographic QCD (IHQCD)~\cite{Gursoy:2007cb, Gursoy:2007er, Gursoy:2008za, Gursoy:2009jd} is a bottom-up 5-D theory inspired by noncritical string theory, that describes the gluon sector of Yang-Mills theories. The model is constructed in such a way to reproduce various features of QCD, for instance, linear confinement, a qualitative hadron spectrum, asymptotic freedom in the UV, and a finite temperature phase diagram that matches SU($N_c$) Yang-Mills theory. Here we will limit ourselves to a pure gluonic sector, but the inclusion of flavor and chiral symmetry breaking is possible by introducing tachyonic D-branes, as well as an axion.
Moreover, we will fit the free parameters of the model comparing with lattice calculations of SU(3) Yang-Mills. We plan to extend our results to other values of $N_c$ in a future publication. In our equations, we will not impose $N_c = 3$ in the expressions which are valid for every $N_c$.
The model consists of a metric $g_{\mu\nu}$ dual to the energy-momentum tensor, the dilaton $\Phi$ dual to ($\lambda_{YM}, \hspace{1 mm} TrF^{2}$) and an axion dual to ($\theta_{YM}, \hspace{1 mm} Tr F \wedge F$ ).
The axion part of the theory can be neglected here, as our interest lies in the thermodynamics mainly. The action for the axion is $N_{c}^{-2}$ suppressed.
In the Einstein frame, the 5-D action which describes this model both at zero and finite temperature is given by 
  \begin{align}\label{eq:S5 action}
    \mathcal{S}_{5} = & -M^{3}_{p}N_{c}^{2} \int d^{5}x \sqrt{g} \left(R -\frac{4}{3} (\partial \Phi)^{2} + V (\Phi)\right) \nonumber \\
    & + 2M_{p}^{3} \int_{\partial\mathcal{M}}d^{4}x\sqrt{h} \mathcal{K}\,,
\end{align}
where $M_{p}$ is the plank mass, $N_{c}$ is the number of colors, $R$ the Ricci scalar, $g$ the metric and $V(\Phi)$ is the dilaton potential. 
The second term in the action is the Gibbons Hawking term that depends on the induced metric $h$ on the boundary, and $\mathcal{K}$ is the extrinsic curvature
\begin{equation}
    K_{\mu\nu} = \nabla_{\mu}n_{\nu} = \frac{1}{2}n^{\rho}\partial_{\rho} h^{\mu\nu} \,\,\,, \mathcal{K} =  h^{ab}K_{ab}\,. 
\end{equation}
Since this is a boundary term it does not affect the solutions to the equations of motion in the zero-$T$ theory, but in the finite $T$ case it plays a pivotal role for providing with a holographically renormalized action~\cite{deHaro:2000vlm,Papadimitriou:2011qb}.
For the dilaton potential we take the ansatz~\cite{Gursoy:2009jd}:
 \begin{equation}\label{eq:V dilaton}
    V = \frac{12}{\ell^{2}}\left(1 + V_{0}\lambda + V_{1}\lambda^{\frac{4}{3}}(\log[1+V_{2}\lambda^{\frac{4}{3}} + V_{3}\lambda^2]^{\frac{1}{2}}) \right)\,,
\end{equation}
where $\lambda = \exp(\Phi)$, and $\ell$ is the AdS length that sets the scale of the fifth dimensional coordinate.
The parameter $V_{0}$ and $V_{2}$ are related to the coefficients of the SU(3) YM $\beta$-function
\begin{equation}\label{eq:V0V2}
V_{0} = \frac{8}{9}b_0\,, \qquad
V_{2} = b_{0}^{4}\left(\frac{23 + 36\frac{b_{1}}{b_{0}^2}}{81 V_{1}} \right)^{2} \,,
\end{equation}
with
%\begin{equation} \label{eq:b0 b1}
%b_0 = \frac{22}{3(4\pi)^2} \quad\text{and}\quad
%\frac{b_{1}}{b_{0}^2}= \frac{51}{121} \,.
%\end{equation}
$b_0 = 22/(3(4\pi)^2)$ and $b_1/b_0^2 = 51/121$.
These values depend on setting $\lambda$ equal to the 't Hooft coupling of the YM theory in the UV. Other normalizations are possible, but do not influence the physical results~\cite{Gursoy:2009jd}.
The free parameters $V_{1},V_{3}$ are set in order to fit lattice results for the thermodynamical properties of the model:~\cite{Gursoy:2009jd}
\begin{equation}
    V_1 = 14\,,\qquad V_3 = 170 \,.
\end{equation}
At finite temperature, after going into imaginary time and compactifying time on a circle $\beta = 1/T$, we identify two types of solutions. The first reads
\begin{equation}\label{eq:metric thermal gas}
ds^{2} = b^{2}_{0}(r)(dr^{2} - dt^{2} + dx^{m}dx_{m}),
\end{equation}
and corresponds to a thermal gas at a temperature $T$. Here $r$ is the coordinate of the fifth dimension, and $b_0^2(r)$ is a scale factor. This is the same metric as in the case of the zero T solution \cite{Gursoy:2007cb,Gursoy:2007er} except for the identification of time being compactifield $t\sim t + i\beta$.
The second solution is a AdS black hole (BH) metric
\begin{equation}\label{eq:BH metric}
    ds^{2} = b^{2}(r)\left(\frac{dr^{2}}{f(r)} - f(r) dt^{2} + dx^{m}dx_{m}\right)\,,
\end{equation}
where the ``blackening'' factor $f(r)$ goes to $0$ at the horizon position $r_h$. Regularity of the solution at the horizon implies that
\begin{equation}\label{eq:T=Th}
    T_h \equiv \frac{|\dot f(r_h)|}{4\pi} = T \,,
\end{equation}
where $T_h$ is the Hawking temperature of the BH.
In the UV ($r\to0$) the two solutions asymptotically coincide, and AdS metric is recovered: $b_0(r)\approx \ell/r$. 
In this setup, the AdS BH metric represents the deconfined phase, and the thermal gas solution corresponds to the confined phase~\cite{Witten:1998zw}.  

For a given temperature $T$, there are either zero or two values of $r_h$ which give a solution with $T_h = T$. These two values identify two separate BH branches, one for small $r_h$, and correspondingly large $b(r_h)$ (big black holes branch), and one at larger $r_h$ and smaller $b(r_h)$ (small black hole branch), which is thermodynamically unstable. Below $T_\mathrm{min}$ there is no BH solution, and the confinement phase transition must complete.
The dilaton profile $\lambda(r)$ grows monotonically from $0$ at $r\to 0$ to $\lambda\to\infty$ at large $r$, and the horizon position $r_h$ corresponds to a finite value $\lambda_h$. A convenient choice is the dilaton frame~\cite{Gubser:2008ny}, in which $\lambda$ is used as the radial coordinate along the fifth dimension.

\begin{figure}[t]
    \def\sepf{0.70}
	\centering
    \includegraphics[height=4.5cm]{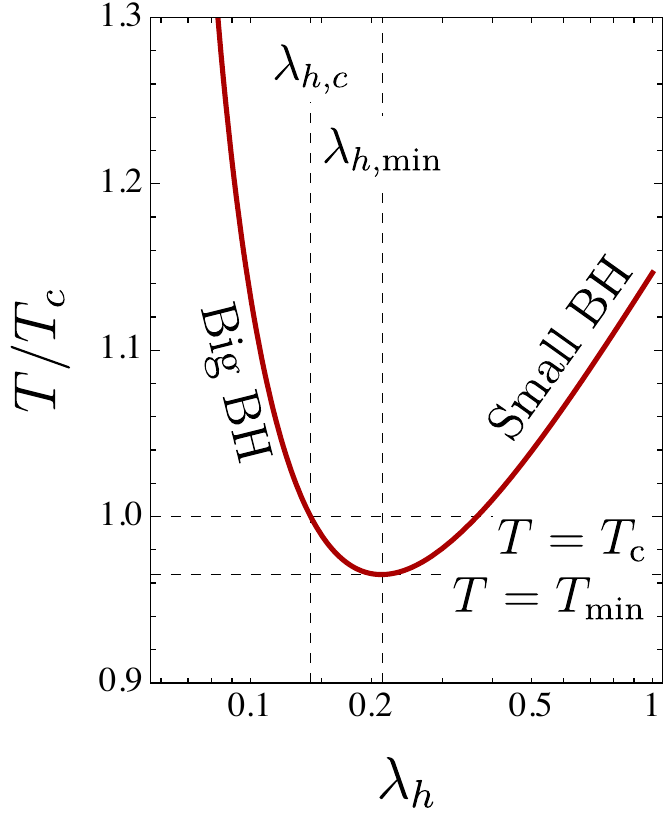}
    \includegraphics[height=4.5cm]{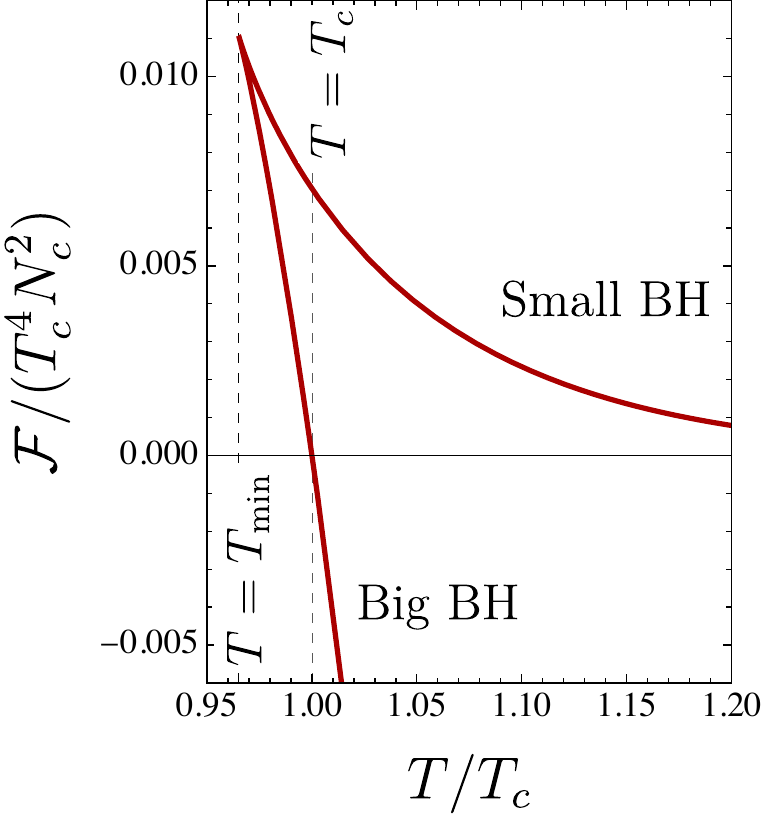}
    \vspace*{-1\baselineskip}
	\caption{Left: temperature as a function of the horizon position $\lambda_{h}$.
	Right: free energy as function of temperature, along the two BH branches.}
	\label{fig:F and T}
\end{figure}

The thermodynamical quantity which controls the phase transition is the free energy difference between the BH solution and the thermal gas one, which is defined as the difference between the actions of the two solutions:
\begin{equation}
    \mathcal{F} = \frac{\beta}{V_3} (\mathcal{S}_\mathrm{dec.} - \mathcal{S}_\mathrm{conf.})\,.
\end{equation}
The action is regularized with a cutoff at $r=\epsilon\to0$, and the difference avoids the need for computing counterterms.
The sign of $\mathcal{F}$ indicates the energetically favorable phase, with $\mathcal{F}<0$ corresponding to the deconfined phase. The critical temperature $T_c$ is defined at $\mathcal{F}=0$.
In practice, the free energy can be computed by integrating the thermodynamic relation $\de\mathcal{F} = - \de S/\de T$ along both black hole branches~\cite{Gursoy:2008za}:
\begin{equation}\label{eq:F integral}
    \mathcal{F} = - \int_{\infty}^{\lambda_h} b(\tilde\lambda_h)^3 \frac{\de T}{\de\tilde\lambda_h}\de\tilde\lambda_h \,.
\end{equation}
Figure~\ref{fig:F and T} shows the temperature and the free energy of the BH solutions.

The entropy is given by the Hawking-Beckenstein formula
\begin{equation}\label{eq:entropy}
    S = \frac{\mathrm{Area}}{4 G_5} = 4\pi M_p^3 N_c^2 V_3 b(r_h)^3 \,,
\end{equation}
where $G_5=1/(16 \pi M_p^3 N_{c}^{2})$ is the 5D Newton constant and $V_3$ is the volume of 3D space.

More details about the geometric properties of the solution can be obtained as discussed in Sec. 7 of Ref.~\cite{Gursoy:2008za} and summarized in the appendix.

%\newpage
\section{Effective Action from Holography}\label{sec:ActionHolography}
% \vspace*{-1\baselineskip}
%\textbf{\smallskip
%\textbf{Effective Action from Holography:}}
In order to study how the deconfinement phase transition took place in the early universe, we need to define an effective action that we can use to compute the transition rate. The qualitative picture is as follows. For $T>T_c$, the free energy is minimized on the BBH solution. In the 4D picture, this corresponds to a deconfined phase. At $T<T_c$, it becomes energetically favorable to tunnel to the free gas solution, corresponding to a confined phase. The phase transition completes before the temperature redshifts below $T_\mathrm{min}$.

In lattice gauge theory or other phenomenological approaches for understanding the confinement phase transition, the conventional order parameter is the vacuum expectation value of the Polyakov Loop. This exhibits a discrete jump at confinement~\cite{Lucini:2012gg,Dumitru:2003hp,Shuryak:2018fjr}.
In our holographic approach, instead, the most natural order parameter is the horizon position $\lambda_h$.
The effective potential for $\lambda_h$ is obtained using a free energy landscape approach, similarly to Refs.~\cite{Creminelli:2001th,vonHarling:2017yew,Baratella:2018pxi, Bigazzi:2020phm} (see also~\cite{Li:2020khm} for an interesting example applied to the Hawking-Page phase transition).
In this approach, one selects a direction in field space that interpolates between the two BH solutions, assuming a specific ansatz for the metric and field configuration, thus reducing the problem of finding the bouncing solution to a single field one~\cite{Creminelli:2001th}. In practice, we proceed as follows. At a given temperature $T$, we construct field and metric configurations as in Eqs.~(\ref{eq:metric thermal gas}),~(\ref{eq:BH metric}), with $\lambda_h$ not restricted to match the BBH or the SBH solutions. These configurations satisfy the equations of motion, except for the condition $T=T_h$, which will be satisfied only for the two values corresponding to the BBH and SBH branches, and violated otherwise. In the latter case, a conical singularity is present at the horizon, and its contribution to the free energy is obtained after regularizing it with a spherical cap (more details are provided in the appendix)~\cite{Solodukhin:1994yz}. The condition posed by the equations of motion guarantee that the chosen field configuration correspond to a local minima of the action with respect to transverse oscillations. Strictly speaking, this does not prove that other tunneling directions do not exist, but we consider it as a good estimate of the bounce action and a lower limit on the transition rate.

\begin{figure}[t]
    \def\sepf{0.70}
	\centering
    \includegraphics[width=.4\textwidth]{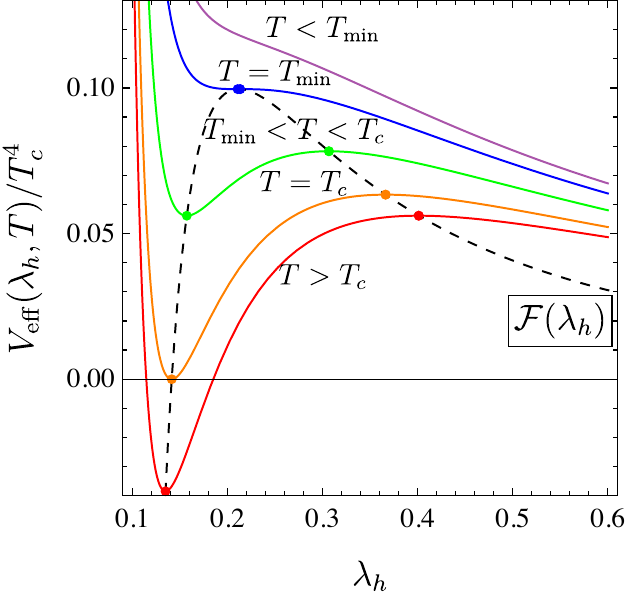} 
     \vspace*{-1\baselineskip}
	\caption{Thermal effective potential as a function of the horizon position $\lambda_{h}$, for different temperatures $T$. The dashed line represents the free energy density of the black hole solution Eq.~(\ref{eq:BH metric}).}
	\label{fig:EffectivePotential}
\end{figure}

We obtain
\begin{equation}\label{eq:Veff}
    V_\mathrm{eff}(\lambda_h, T) = \mathcal{F}(\lambda_h) - 4\pi M_p^3 N_c^2 b(\lambda_h)^3 \left(1- \frac{T_h}{T}\right) \,.
\end{equation}
Here $\mathcal{F}$ is computed using Eq.~(\ref{eq:F integral}) with $T_h$ in the integral. Even though $T\neq T_h$, this relation can be used to compute the action of a given field configuration. The same result can be obtained from the UV asymptotics of $b(\lambda), f(\lambda)$.
The result is shown in Fig.~\ref{fig:EffectivePotential}. We see that the potential reproduces the expected features from the discussion above. For $T>T_\mathrm{min}$ the potential has a minimum corresponding to the BBH solution, a maximum corresponding to the unstable SBH, and another minimum at $\lambda_h\to \infty$, where the free gas solution is recovered. Below $T_\mathrm{min}$, the latter is the only extremum.

The tunneling proceeds through the nucleation of a bubble that interpolates between the BBH solution at infinity and some unstable, singular configuration at the center, rapidly decaying to the confined thermal gas phase. The bounce solution goes through the unstable SBH solution. This is the equivalent, in our setup, of the Hawking-Page transition in 4D space-time, in which the SBH solution acts as an instanton connecting the BBH solution to AdS space-time~\cite{Hawking:1982dh}.

The other ingredient that we need in order to define an effective action is the kinetic term. In principle, this can be computed by evaluating the dilaton kinetic term and the Ricci scalar term in Eq.~(\ref{eq:S5 action}) on a configuration as discussed above, and extracting the term proportional to $(\vec\nabla\lambda_h(\vec x))^2$, where $\vec x$ and $\nabla x$ are the 3-space coordinates and spatial derivatives.
This is anyway a complicated task, which we postpone for a future investigation.
Here, we will assume a kinetic term~\cite{Baratella:2018pxi}
\begin{equation}\label{eq:kinetic term}
    c \frac{N_c^2}{16\pi^2}(\vec\nabla\lambda_h)^2\,,
\end{equation}
and we vary $c$ in the range, $0.3 - 3$.
The dependence on the kinetic term appears to be moderate, in contrast to what happens in the regime of strong supercooling~\cite{Baldes:2021aph}.
The bounce action is the sum of Eqs.~(\ref{eq:Veff}) and (\ref{eq:kinetic term}), computed on the bounce solution:
\begin{equation}\label{eq:EffectiveAction}
    \mathcal{S}_B = \frac{4\pi}{T}\int \de r \, r^2 \left[ c \frac{N_c^2}{16\pi^2}(\partial_r\lambda_h(r))^2 + V_\mathrm{eff}(\lambda_h(r), T)\right]
\end{equation}
where we assumed an $\mathcal{O}(3)$ symmetric action, as we are interested in thermal tunneling. The bounce solution is obtained using the shooting method with boundary conditions $\lambda_h(r\to\infty) = \lambda_h^\mathrm{BBH}$ and $\partial_r\lambda_h(r)|_{r=0} = 0$. We double-checked our results using the publicly available code FindBounce~\cite{Guada:2020xnz}. The tunneling rate per unit volume and time is then~\cite{Linde:1981zj}
\begin{equation} \label{eq:NucleationRate}
    \Gamma = T^4 \left(\frac{\mathcal{S}_B}{2\pi}\right)^{3/2} e^{-\mathcal{S}_B} \,.
\end{equation}

\section{Gravitational Waves}
% \vspace*{-1\baselineskip}
%\label{sec:GWs}
%\smallskip
%\textbf{Gravitational Waves:}
We now turn our attention to the calculation of the GW spectrum. The phase transition parameters that we need for this estimate are the inverse duration of the phase transition in units of the Hubble time $\beta/H$, the energy liberated during the phase transition $\alpha$, the kinetic energy of the bubble walls $\kappa(\alpha)$, and finally the bubble wall velocity $v_w$. In this work, we only concern ourselves with the sound wave contribution to the gravitational wave spectrum~\cite{Hindmarsh:2015qta,Hindmarsh:2019phv}, also including the suppression of having a short lived source \cite{Guo:2020grp,Ellis:2020awk,Ellis:2019oqb,Gowling:2021gcy}. We will employ the templates presented in the works of the LISA Cosmology Working Group \cite{Caprini:2019egz,Caprini:2015zlo}.

In order to compute the GW spectrum, we need to specify the energy scale of the theory. Pure Yang-Mills theories are determined by a single energy scale, with all other scales being proportional to this one. Equivalent choices may be the mass of the lightest glueball, the strong coupling scale $\Lambda$ (defined as the energy at which the perturbative coupling constant formally diverges), or the critical temperature $T_c$. In SU(3) IHQCD, one finds $m_0 \approx 5.1 \Lambda$ and $m_0 \approx 6 T_c$, in quite good agreement with lattice results~\cite{Gursoy:2007er, Gursoy:2009jd}. In the following, we will use $T_c$ to set the energy scale, as it will be very natural to compare it with the nucleation and percolation temperatures.

We start by estimating the nucleation and percolation temperatures as functions of the critical temperature $T_c$.
The nucleation temperature $T_n$ is determined by the condition that the nucleation rate per Hubble volume per Hubble time equals 1, i.e., $\Gamma(T)/ H^{4} = 1$. For $c=1$, we obtain $T_n = 0.992 T_c$ for $T_c = 100\GeV$ and $T_n = 0.993 T_c$ for $T_c = 50\MeV$.
Having $T_n$ close to $T_c$ is not unexpected, as it could be estimated from thermodynamics using some lattice input (see e.g. Ref.~\cite{GarciaGarcia:2015fol}).

The percolation temperature roughly indicates the end of the phase transition. It is defined as the time when the probability $\mathcal{P}$ of remaining in the false vacuum is reduced by $\mathcal{O}(30)\%$
\begin{align}\label{eq:PercTemp}
    \mathcal{P}(t) &= e^{-I(t_p)},   \\
    I(t) & = \frac{4\pi}{3} \int_{t_c}^t dt' \Gamma(t') a(t')^3 r(t,t')^3 \,,
    % I(T_{p}) &= \frac{4\pi}{3}\int_{T}^{T_{c}} \frac{dT' \Gamma(T')}{H(T')T'^4} \left( \int_{T}^{T'} dT'' \frac{v_{w}(T'')}{H(T'')} \right)^3
\end{align}
where
\begin{equation}
    r(t,t') = \int_{t'}^t dt'' \frac{v_w}{a(t'')}
\end{equation}
is the radius at time $t$ of a bubble emitted at $t'$.
The precise definition of $T_p$ varies across the literature. Here we impose $I(T_p) = 0.34$ as discussed e.g. in Ref.~\cite{Ellis:2018mja}. An alternative definition, leading to smaller $T_p$, is $I(T_p) = 1$~\cite{Croon:2020cgk, Enqvist:1991xw}.
We assume a constant value of $v_w$, and we obtain $T_p = (0.993\pm 0.003)T_c$ for $v_w = 0.01 - 1$ and for both $T_c = 50\MeV$ and $100\GeV$, where the uncertainty in $T_p$ comes from varying $c$ and only negligibly from $v_w$. We will discuss the wall velocity in more detail below. We see that the ratios $T_n/T_c$, $T_p/T_c$ are almost independent of the critical temperature $T_c$. This is due to the strong exponential dependence of $\Gamma(T)$ on $T/T_c$.

\begin{figure}[t]
    \def\sepf{0.65}
	\centering
    \includegraphics[scale=\sepf]{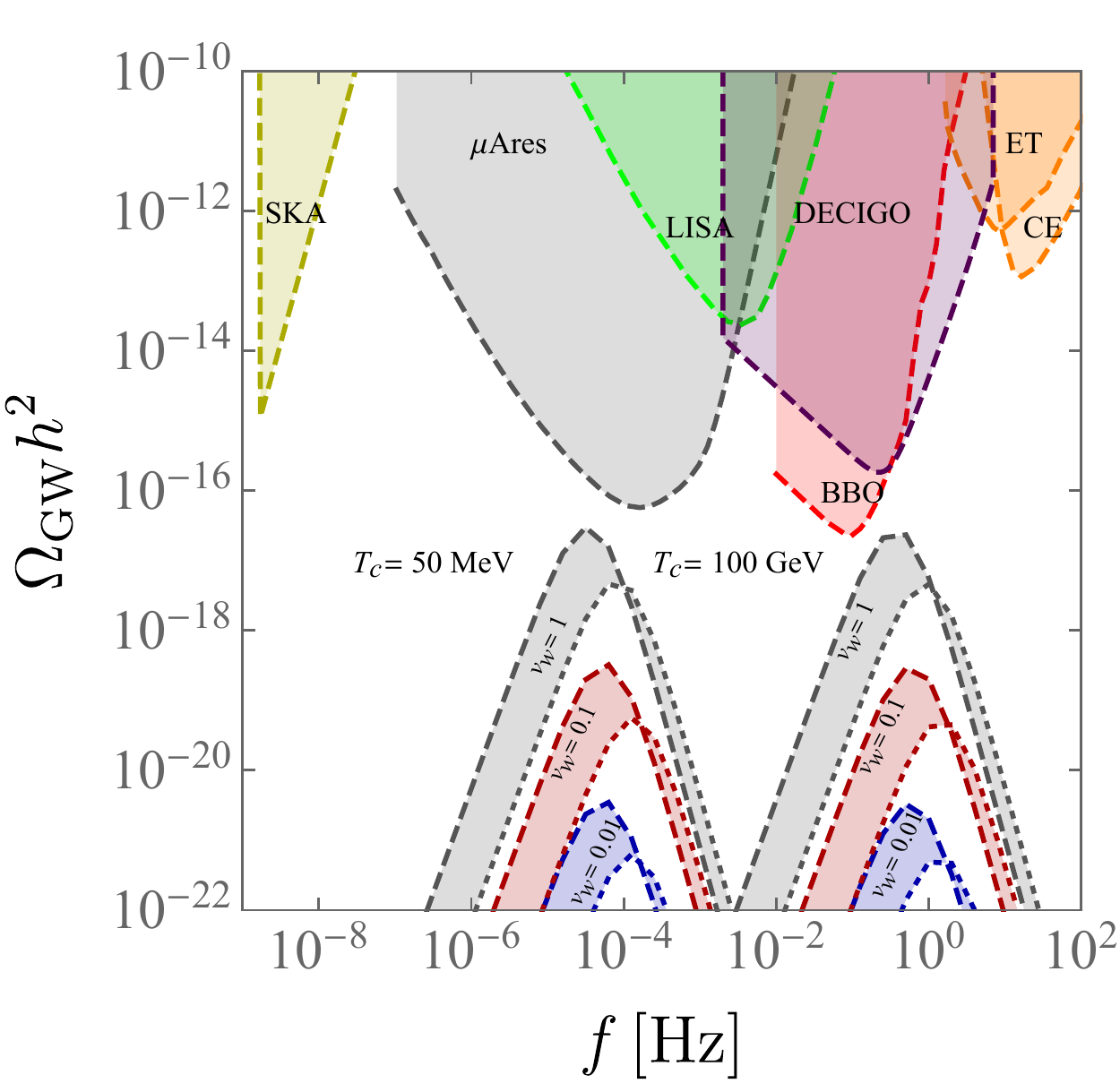} 
	\caption{Gravitational wave spectra estimated with our effective action for IHQCD and the projected sensitivity curves for future GW experiments: Square Kilometer Array (SKA)~\cite{Janssen:2014dka}, $\mu$Ares~\cite{Sesana:2019vho}, LISA~\cite{LISA:2017pwj}, DECIGO/BBO~\cite{Yagi:2011wg}, Einstein Telescope (ET)~\cite{Punturo:2010zz}, and Cosmic Explorer (CE)~\cite{LIGOScientific:2016wof}. For illustration, we choose a critical temperature $T_{c}=50\MeV$ and $T_c=100\GeV$, and the contours denote $v_w=1$ (gray), $v_{w}=0.1$ (red) and $v_{w}=0.01$ (blue).}
	\label{fig:Gravitationalwaves}
\end{figure}

The parameter $\beta/H$ describes the duration and the number of nucleated bubbles the phase transition generates, and is evaluated when the phase transition has completed, i.e. at the percolation temperature $T_p$. For a fast phase transition one can approximate $\Gamma\sim\exp[\beta(t-t_p)]$, and the inverse duration of the phase transition is given by
\begin{equation}\label{eq:beta}
    \frac{\beta}{H} = T\left( \frac{d\mathcal{S}_{B}}{dT}\right)\bigg|_{T = T_{p}}\,.
\end{equation} 
We obtain $\beta/H\sim\mathcal{O}(10^5)$ (the exact values are summarized in Tab.~\ref{tab:GWparameters}), with an uncertainty of order $10\%$ stemming from $v_w$, while the uncertainty from varying $c=0.3-3$ is of order $\mathcal{O}(1)$ and is indicated by the width of the bands in our GW spectra in Fig.~\ref{fig:Gravitationalwaves}.

The next quantity that we need to compute is the strength of the phase transition $\alpha$, i.e. the amount of energy released during the phase transition that is available to convert into the fluid motion of the plasma. We define it as
\begin{equation}
    \alpha = \frac{4}{3}\frac{\Delta \theta}{\Delta w} = \frac{1}{3} \frac{\Delta \rho - 3 \Delta p}{\Delta w}.
\end{equation}
where $\theta$ is the trace of the energy-momentum tensor, $w$ is the enthalpy, and $\Delta$ indicates that we take the difference of the corresponding values in the deconfined and confined phases. 

The enthalpy and trace anomaly are given by $\Delta w = T\cdot \Delta s$ and $\Delta\theta = 4\mathcal{F} + T\cdot  \Delta s$. 
We obtain $\alpha|_{T_p} \approx 0.343$, with a $\mathcal{O}(10^{-2})$ relative uncertainty coming from the variation in $c$ in the evaluation of $T_p$, and an even smaller dependence on $T_c$ and on $v_w$. Our result differs from the lattice one of Ref.~\cite{Caselle:2018kap} by roughly 10\%, which we consider to be a good estimate of the overall uncertainty on $\alpha$.

The calculation of wall velocity $v_w$ in cosmological phase transitions has received a lot of attention throughout the years. An estimate of $v_w$ is typically obtained by computing the transmission coefficient of particles at the bubble wall~\cite{Bodeker:2009qy, Bodeker:2017cim, Konstandin:2014zta, Megevand:2009gh, Hoche:2020ysm, Wang:2020zlf,Azatov:2020ufh, Dorsch:2021ubz, Gouttenoire:2021kjv}, or can be understood from the local thermodynamics properties of the plasma~\cite{Ai:2021kak, Balaji:2020yrx}.
In strongly coupled theories the problem becomes even more complicated, and can be addressed using holography in certain models~\cite{Baldes:2020kam, Bigazzi:2021ucw, Bea:2021zsu, Bea:2022mfb, Janik:2022wsx}.

Extrapolating the result of Refs.~\cite{Bigazzi:2021ucw, Bea:2021zsu, Janik:2022wsx} to our parameter range, we obtain $v_w\sim \mathcal{O}(0.01)$. Even smaller velocities are obtained in Ref.~\cite{Asadi:2021pwo}. On the other hand, Ref.~\cite{Bea:2022mfb} obtains a terminal bubble wall velocity of $v_w \sim 0.3$ in a 3+1 dimensional simulation of the bubble growth in a regime of at least moderately strong supercooling. Finally if one resorts to the Chapman-Jouguet formula for the wall velocity we obtain $v_\mathrm{CJ}\approx 0.867$.
Under these circumstances, we treat the bubble wall velocity as a free parameter and leave it for future work.

Figure~\ref{fig:Gravitationalwaves} shows our results for the GW spectra, together with the expected sensitivity of future observatories. The contours are evaluated by computing the effective action Eq.~(\ref{eq:EffectiveAction}), varying $c=0.3-3$. The dashed line corresponds to $c=3$, the dotted to $c=0.3$, with $c=1$ in between. The variation of $c$ affects the GW spectrum mainly through $\beta/H$. 

\begin{table}[t]
    \centering
    \begin{tabular}{|c|c|c|c|c|}
      \hline
          & $\alpha$ & $\beta/H\, (v_{w}=1)$ & $\beta/H\, (0.1)$ & $\beta/H\, (0.01)$\\
         \hline
        $T_{c}= 50 \MeV$ & 0.343 & 9.0 $\times 10^4$ & $8.6\times 10^4$ & $8.2 \times 10^4$\\
         \hline
        $100 \GeV$ & 0.343 & $6.8\times 10^4$ & $6.4\times 10^4$ & $6.1\times 10^4$\\
         \hline
    \end{tabular}
    \caption{Values of $\beta/H$ and $\alpha$ for different wall velocities and critical temperatures. All entities are evaluated at the percolation temperature $T_{p}= 0.993T_{c}$.}
    \label{tab:GWparameters}
\end{table}

%\newpage
\section{Discussion}
%\label{sec:Discussion}
%\smallskip
%\textbf{Discussion:}
In this work, we report on the first computation of the GW signal from the confinement/deconfinement phase transition in pure SU(3) Yang-Mills theory using bottom-up holography. We use the IHQCD framework which successfully reproduces lattice results for the equilibrium thermodynamics of this theory, and calculate the equilibrium and quasiequilibrium quantities of the phase transition relevant for GWs. 
These are the energy budget $\alpha$, the percolation temperature $T_{p}$ and the average bubble size compared to the Hubble horizon  $\beta/H$, which we obtain with $\mathcal{O}(1)$ errors.
Our calculation of $\beta/H$ agrees up to $\mathcal{O}(1)$ with previous estimates based on effective models of low energy QCD~\cite{Halverson:2020xpg,Huang:2020mso,Kang:2021epo}. 

The recent works of Refs.~\cite{Ares:2021ntv, Ares:2020lbt, Ares:2021nap} also employ holographic techniques for studying phase transition dynamics and the resulting GWs, however their holographic models do not aim to quantitatively reproduce the behavior of known strongly coupled theories. References~\cite{Bigazzi:2020avc, Bigazzi:2020phm} study the WSS model, which can reproduce qualitative features of QCD. References~\cite{Ahmadvand:2017xrw,Li:2021qer,Cai:2022omk} also use holography to model the phase transition of QCD-like theories, however they do not calculate $\beta/H$ and instead choose an optimistic value. Our study suggests that their GW signal predictions are grossly overestimated because of this. 

The resulting GW spectra are shown in Fig.~\ref{fig:Gravitationalwaves}. Even for the most optimistic case of highly relativistic bubble walls, the signal is out of reach for next generation GW detectors. However, we expect a magnification of the GW signals for larger $N_{c}$ due to additional supercooling from delaying nucleation by having additional degrees of freedom. We intend to elaborate on this in future work by utilizing the methods presented here for SU(3) case to the SU($N_{c}$) case.
Additional questions left for future work are the inclusion of flavor to study chiral symmetry breaking/confinement, the glueball spectra for $N_c>3$ and the impact of an axion on the deconfinement temperature.

\smallskip
\textbf{Acknowledgments:} We thank Mark Hindmarsh, Manuel Reichert and Michele Redi for interesting discussion about holography and phase transitions. We also thank James Renwick Hudspith for valuable discussion on $SU(N_c)$ lattice thermodynamics. 
Work in Mainz is supported by the Deutsche Forschungsgemeinschaft (DFG), Project No. 438947057 and by the Cluster of Excellence ``Precision Physics, Fundamental Interactions, and Structure of Matter'' (PRISMA+ EXC 2118/1) funded by the German Research Foundation (DFG) within the German Excellence Strategy (Project No. 39083149).

\newpage

%%%%%%%%%%%%%%%%%%%%%%%%%%%%%%%%%%%%%%%%%%%%%%%%%%%%%%%%%%%%%%%%
%%%%%%%%%%%%%%%%%%%%%%%%%%%%%%%%%%%%%%%%%%%%%%%%%%%%%%%%%%%%%%%%
%%%%%%%%%%%%%%%%%%%%%%%%%%%%%%%%%%%%%%%%%%%%%%%%%%%%%%%%%%%%%%%%

\section*{Appendix}

\subsection*{Equations of motion and thermodynamics of IHQCD}

Starting from the black hole ansatz of Eq.~(\ref{eq:BH metric}), the Einstein equations are written as
\begin{align}
& 6\frac{\dot{b}^{2}}{b^{2}} + 3\frac{\Ddot{b}^{2}}{b^{2}} + \frac{\dot{b}}{b}\frac{\dot{f}}{f} = \frac{b^{2}}{f} V \\
& 6\frac{\dot{b}^{2}}{b^{2}} - 3\frac{\Ddot{b}^{2}}{b^{2}} = \frac{4}{3}\dot{\Phi}^{2} 
\\
& \frac{\Ddot{f}}{f} + 3\frac{\dot{b}}{b} = 0
\\
& \Ddot{\Phi} + \left(\frac{\dot{f}}{f} + 3\frac{\dot{b}}{b} \right)\dot{\Phi} + \frac{3b^{2}}{8f}\frac{dV}{d\Phi} 0
\end{align}
where the dilaton potential is given in Eq.~(\ref{eq:V dilaton}). The thermal gas equations of motion are obtained by setting $f(r)= 0$ and replacing $b(r) \rightarrow b_0(r)$. A very convenient way of solving the system is to define two scalar variables
\begin{equation}
    X(\Phi) = \frac{\Phi'}{3A'}\,, \qquad Y(\Phi) = \frac{g'}{4 A'} \,,
\end{equation}
where we have defined $A = \log b$, $g = \log f$ and with a prime we denote a derivative with respect to the coordinate of the fifth dimension in the domain-wall frame, $u$ defined such that $\de u = e^A \de r$. Expressing all these quantities as functions of $\Phi$ (or, equivalently, $\lambda = e^\Phi$), the Einstein equations turn into first-order differential equations for $X,Y$:
\begin{align}
    \frac{\de X}{\de\Phi} & = -\frac{4}{3}(1 - X^2 + Y) \left(1 + \frac{3}{8X}\frac{\de\log V}{\de\Phi}\right) \\
    \frac{\de Y}{\de\Phi} & = -\frac{4}{3}(1 - X^2 + Y) \frac{Y}{X} \,,
\end{align}
where the initial conditions are set close to the horizon and descend from the requirement of regularity only:
\begin{align}
    Y(\Phi) & \approx \frac{Y_h}{\Phi_h - \Phi} + Y_1 \label{eq:eom Y} \\
    X(\Phi) & \approx -\frac{4}{3}Y_h + X_1(\Phi_h - \Phi) \label{eq:eom X} \,,
\end{align}
with
\begin{align}
    Y_h = \frac{9}{32} \frac{V'(\Phi_h)}{V(\Phi_h)}
\end{align}
and
\begin{align}
    X_1 & = \frac{3}{16}\left( \frac{V''(\Phi_h)}{V(\Phi_h)} - \frac{V'(\Phi_h)^2}{V(\Phi_h)^2} \right) \\
    Y_1 & = \frac{9}{64}\left( \frac{V''(\Phi_h)}{V(\Phi_h)} - 2\frac{V'(\Phi_h)^2}{V(\Phi_h)^2} \right) -1 \,.
\end{align}
The only free parameter is the position of the horizon in the dilaton frame, $\Phi_h$ (or, equivalently, $\lambda_h$). Equation~(\ref{eq:eom X}), with $Y=0$, describes instead the thermal gas solution, with initial condition
\begin{equation}
    X_0(\Phi\to\infty) = -\frac{1}{2} - \frac{3}{16\Phi} \,.
\end{equation}
The geometry of the solution can be obtained from the knowledge of $X,Y$:
\begin{align}
    A & = A_0 + \int_{\Phi_0}^\Phi \frac{\de\tilde\Phi}{3X} \label{eq:A X}\\
    g & = \int_{\Phi_0}^\Phi \frac{4}{3}\frac{Y}{X}\de\tilde\Phi \label{eq:g X} \,,
\end{align}
where $e^{A_0}$ is the value of the scale factor at position $\Phi_0$. Clearly, there is some redundancy, in that one can vary $\Phi_0$ and $A_0$ in such a way to obtain the same theory. The only physical quantity is the scale $\Lambda$, which enters the UV behavior of the scale factor and of $\lambda$. This is defined from the scale factor at some small value $\lambda_0$ as
\begin{equation}
    \Lambda \ell \equiv \exp \left[ A(\lambda_0) - \frac{1}{b_0 \lambda_0}\right] (b_0 \lambda_0)^{-b_1/b_0^2} \,,
\end{equation}
where the RHS is constant for $\lambda_0 \to 0$. This scale is the perturbative strong coupling scale, appearing in the UV expansion of the coupling
\begin{equation}
    \frac{1}{b_0 \lambda} = \log\frac{E}{\Lambda_p} - \frac{b_1}{b_0^2} \log\log \frac{E}{\Lambda} + \dots \,,
\end{equation}
where we identified the energy $E$ on the 4D field theory side with the 5D scale factor $b(r) = e^{A(r)}$.
Correspondingly, $\Lambda$ controls the UV behavior of the solutions close to the UV boundary at $r\to 0$, where the geometry asymptotes to AdS:
\begin{align}
    b(r) & = \frac{\ell}{r}\left[ 1 + \frac{4}{9}\frac{1}{\log r\Lambda} - \frac{4}{9}\frac{b_1}{b_0^2}\frac{\log(-\log r\Lambda)}{(\log r\Lambda)^2} + \dots\right] \,,\\
    b_0 \lambda(r) & = -\frac{1}{\log r\Lambda} + \frac{b_1}{b_0^2}\frac{\log(-\log r\Lambda)}{(\log r\Lambda)^2} + \dots
\end{align}
The scale $\Lambda$ can be used to set the energy scale of the theory. Alternatively, one can fix the energy scale by choosing the mass of the lightest glueball or the critical temperature of the confinement phase transition, as we do in this paper.

In principle, Eqs.~(\ref{eq:A X}), (\ref{eq:g X}), together with Eqs.~(\ref{eq:T=Th}), (\ref{eq:entropy}) and (\ref{eq:F integral}) are enough to compute all the thermal properties of the system. Nevertheless, it is interesting to notice that the same quantities can be extracted from the UV behavior of $X,Y$. In the UV limit $\lambda\to0$, $X,Y$ satisfy
\begin{equation}
    Y(\lambda) = \mathbf{Y_0} e^{-\frac{4}{b_0 \lambda}} (b_0 \lambda)^{-4b_1/b_0^2} 
\end{equation}
\begin{align}
    X(\lambda) - & X_0(\lambda) = \nonumber \\
    & \left(\frac{\mathbf{Y_0}/2 - \mathbf{C_0}}{X_0(\lambda)} + \mathbf{C_0} X_0(\lambda) \right) e^{-\frac{4}{b_0 \lambda}} (b_0 \lambda)^{-4b_1/b_0^2} \,,
\end{align}
where $b_0,b_1$ are defined in the main text. Here $\mathbf{C_0},\mathbf{Y_0}$ are constants that determine, respectively, the energy density and the entropy of the system, while the combination $\mathbf{C_0}-\mathbf{Y_0}/2$ determines the vev of the gluon condensate.
Indeed, one can show that the following relations hold:
\begin{align}
    T & = \frac{\Lambda^4 \ell^3}{\pi}\frac{\mathbf{Y_0}(\lambda_h)}{b(\lambda_h)^3}\label{eq:AppTemp} \\
    S T & = 4 M_p^3 N_c^2 V_3 \mathbf{Y_0} \ell^3 \Lambda^4 \label{eq:AppEntropy}\\
    \mathcal{F} & = M_p^3 N_c^2 \Lambda^4 \ell^3 (6\mathbf{C_0} - 4\mathbf{Y_0})\label{eq:AppFree}\,.
\end{align}
Moreover, the energy density, the specific heat, and the speed of sound of the system are given by
\begin{align}
    \rho & = 6 M_p^3 N_c^2 \mathbf{C_0} \Lambda^4 \ell^3 \\
    c_v & = 6 M_p^3 N_c^2 \Lambda^4 \ell^3 \frac{\de \mathbf{C_0}}{\de T} \\
    c_s^2 & = \left( \frac{\de\log\mathbf{Y_0}}{\de\log T} - 1 \right)^{-1} \,.
\end{align}
Finally, the constants $\mathbf{Y_0},\mathbf{C_0}$ are related by
\begin{equation}
    \mathbf{C_0}(T) = \frac{2}{3}\mathbf{Y_0}(T) -\frac{2}{3}\int_{T_c}^T \mathbf{Y_0}(\tilde T) \de\log\tilde T \,.
\end{equation}
We checked numerically that the results using $\mathbf{Y_0},\mathbf{C_0}$ are consistent with the ones obtained from Eqs.~(\ref{eq:A X}), (\ref{eq:g X}).

\subsection*{IHQCD comparison with $\mathbf{SU(N_c)}$ lattice results}\label{sec:lattice}
Given our numerical solutions for the thermodynamical quantities $(T, \mathcal{F}, S)$, we can calculate the energy density $\rho$, pressure $P$ and $\theta = \rho- 3P$ and compare them to lattice results. 

 The authors of~\cite{Gursoy:2009jd} used the lattice data of~\cite{Boyd:1996bx} and achieved a qualitatively good fit for the entropy density, energy density, and pressure with the parameters of the dilaton potential $V1= 14 \,, V3=170$. The work by~\cite{Panero:2009tv} performs a lattice study for the equilibrium thermodynamical properties of $SU(N_{c})$ theories varying $N_{c}=3-8$ and showed that IhQCD provides an overall good fit to the various gauge groups. 
 
\begin{figure}[t]
    \def\sepf{0.65}
	\centering
    \includegraphics[scale=\sepf]{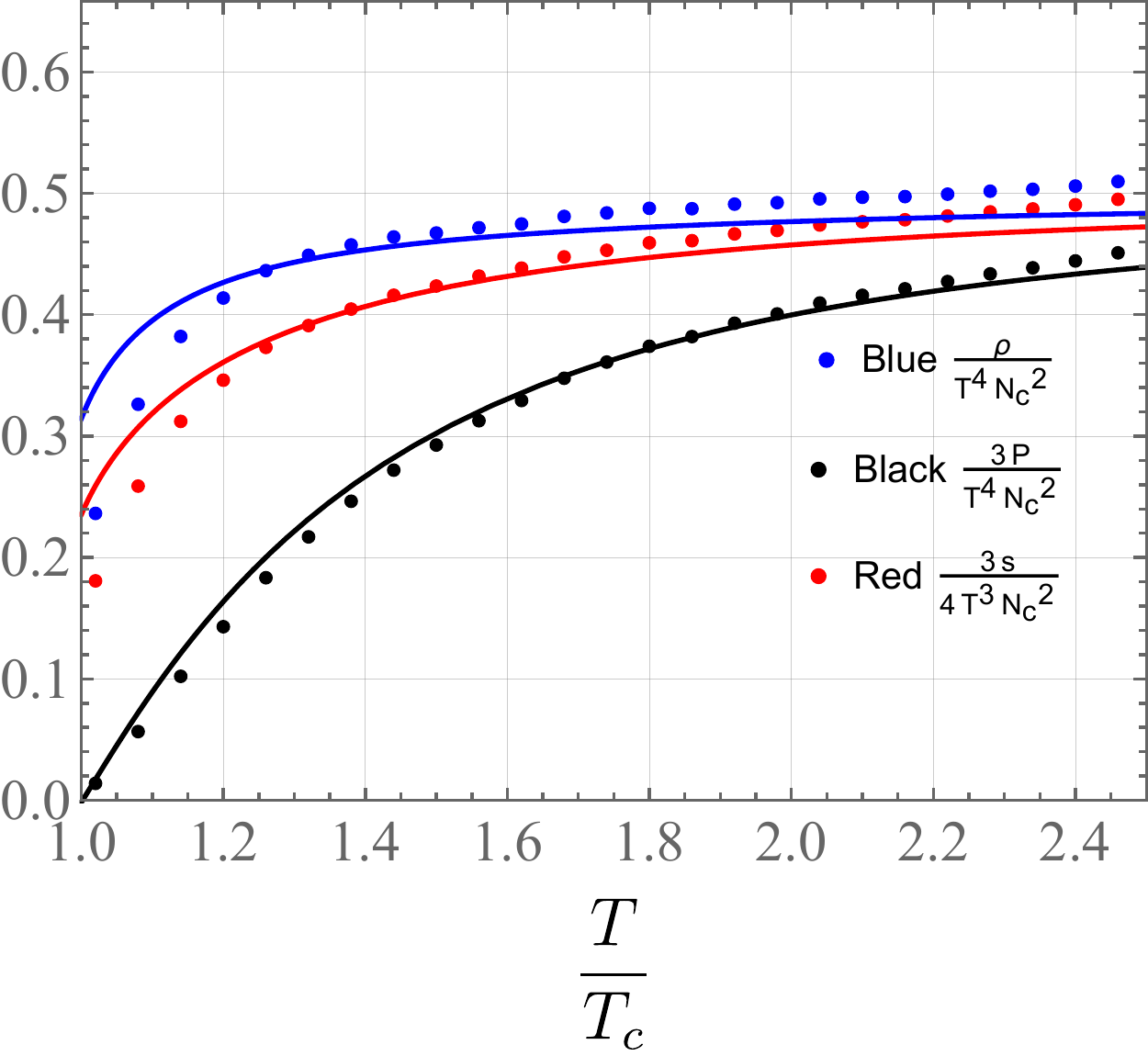} 
	\caption{SU(3) equilibrium thermodynamics from \cite{Caselle:2018kap} in dotted points and where the lines are the prediction from IhQCD for the dilaton potential parameters $V_1 = 14\, \hspace{1mm} V_3 = 170$. The y-axis ends at the SB-limit for the thermal gas limit of $\frac{\pi^2}{15}$.}
	\label{fig:Lattice}
\end{figure}

 Lattice observables of equilibrium thermodynamical quantities are extracted mainly in the continuum limit, where the error induced by the lattice spacing gets minimized. Here we reproduce the results in~\cite{Gursoy:2009jd} using the latest lattice data from ~\cite{Caselle:2018kap} in the case of $SU(3)$ and fit to the data, see Fig.~\ref{fig:Lattice}.
 We find that
 the same parameters values as in~\cite{Gursoy:2009jd} provide an adequate fit to the newer data. 
 For a thorough introduction to lattice gauge theory and how their observables are constructed we refer to Refs.~\cite{Greensite:2003bk, Brambilla:2014jmp, Kogut:1982ds, Shuryak:2018fjr}.

We use Eq.~(7.38) in \cite{Gursoy:2008za} to obtain the first lattice observable, i.e. the entropy density $s/T^{3}$ as a function of the dilaton potential and the scalar variable $X(\lambda)$. To find the pressure and energy density we compute the free energy $\mathcal{F}$, and use the thermodynamical relations i.e. $P=-\mathcal{F}, \hspace{1 mm} \rho = \mathcal{F} + ST$ and Eqs. (\ref{eq:AppTemp}), (\ref{eq:AppEntropy}), (\ref{eq:AppFree}) to obtain the lattice observables provided the proper normalization. The results of our fit to the lattice data are shown in Fig. 4, together with the extracted best-fit parameters for $V_1 \,, V_3$.

\subsection*{Effective potential}

Here we show the details of the calculation of the effective potential.
We start from the Euclidean BH metric
\begin{equation}\label{eq:BH metric appendix}
    ds^{2} = b^{2}(r)\left(\frac{dr^{2}}{f(r)} + f(r) dt^{2} + dx^{m}dx_{m}\right)\,,
\end{equation}
which we expand in the vicinity of $r_h$ using
\begin{equation}
    f(r) \approx \dot f_h(r-r_h) \,, \qquad b(r) \approx b_h \,,
\end{equation}
obtaining
\begin{equation}
    ds^2 = b_h^2\left(\frac{dr^{2}}{\dot f_h(r-r_h)} + \dot f_h(r-r_h) dt^{2} + (d\vec x)^2\right)\,.
\end{equation}
Let us now define
\begin{align}
    y & = \frac{2b_h}{(\dot f_h)^{1/2}}\sqrt{r-r_h} \qquad \text{with } y>0 \,, \\
    \varphi & = 2\pi T t \qquad \text{with } 0<\varphi<2\pi \,, 
\end{align}
The metric (restricting ourselves to the $r,t$ directions) is
\begin{equation}
    ds^2 = dy^2 + y^2 \left(\frac{\dot f_h}{4\pi T}\right)^2 d\varphi^2
\end{equation}
which is the metric of a cone, where $y$ is the distance from the tip, $\varphi$ the angular direction and the angle $\alpha$ of the cone is
\begin{equation}
    \sin\alpha = \frac{\dot f_h}{4\pi T} = \frac{T_h}{T}
\end{equation}
The singularity, which disappears for $\sin\alpha=1$, can be regularized by cutting the cone surface at some small $y_s$, and smoothly gluing a spherical cap of radius $R_s = y_s \tan\alpha$.
In polar coordinates, the metric on the spherical cap is $(R_s^2,R_s^2\sin\theta^2,b(y)^2,b(y)^2,b(y)^2)$, with $0<\theta<\pi/2-\alpha$. We can now compute the contribution of the cap to the action:
\begin{equation}
    \mathcal{S}_\mathrm{cone}^{BH} = -M_p^3 N_c^2 \int d^5 x \sqrt{g} [\mathcal{R}-\frac{4}{3}(\partial\phi)^2 + V(\phi)]
\end{equation}
The matter contribution to the action is regular and goes to zero as the cap is shrunk.
The geometric part is the relevant one, and it is easily evaluated using $\sqrt{g}= R_s^2\sin\theta b(y)^3\approx R_s^2\sin\theta b_h^3$, the area of the spherical cap is $2\pi R_s^2 (1-\sin\alpha)$
and $\mathcal{R}= 2/R_s^2 + \dots$ where the dots are terms containing the derivatives $\partial_\theta b\propto R_s \to 0$. The result is thus
\begin{align}
    \mathcal{S}_\mathrm{cone,\,geom}^{BH} & = -M_p^3 N_c^2 \int d^5 x \sqrt{g} \mathcal{R} \nonumber\\
%    & = -M_p^3 N_c^2 V_3 b_h^3 \mathrm{Area} \frac{2}{R_s^2} \nonumber\\
    &= -4\pi M_p^3 N_c^2 V_3 b_h^3 (1-T_h/T) \,.
\end{align}

\subsection*{LISA Cosmology Working Group Gravitational Waves Template}\label{sec:LisaTemplate}

In the context of gravitational waves emitted from FOPT's numerous numerical studies \cite{Caprini:2019egz, Caprini:2015zlo, Hindmarsh:2017gnf,  Hindmarsh:2019phv, Cutting:2019zws} provides the literature with an estimated GW spectrum based on knowledge of the parameters $\alpha,\hspace{1mm}\beta/H,\hspace{1mm} v_{w},\hspace{1mm} g_{*},\hspace{1mm} T_{*}$. In this work, we use the estimated formula provided by the LISA cosmology working group \cite{Caprini:2019egz} \begin{equation}
\label{eq:LISA GW simulation}
\frac{d \Omega_{GW,0}}{d \ln{f}} = 0.687F_{GW,0} K^{\frac{3}{2}}\left(H(T_{*})R(T_{*})\right)^2\Tilde{\Omega}_{GW}C\left(\frac{f}{f_{p,0}}\right)\,, 
\end{equation} where the prefactor $F_{GW,0} = (3.57 \pm 0.05)\cdot 10^{-5}
\left(\frac{100}{g_{*}}\right)^{\frac{1}{3}}
$ accounts for the redshift. The expression for $K$ describes the fraction of kinetic energy available during the transition \begin{equation}
        K = \frac{\kappa(\alpha) \alpha}{1 + \alpha},
    \end{equation} where \begin{equation}
    \kappa(\alpha) = \frac{\alpha}{ 0.73 + 0.083\sqrt{\alpha} + \alpha}\,,
\end{equation} is the efficiency factor for wall speeds $v_w \approx 1$, which gets modified for lower wall velocities~ \cite{Espinosa:2010hh}.
The scaling of $K$ in Eq.~(\ref{eq:LISA GW simulation}) indicates that we in this work have a short lived source which induces a suppression to the spectrum of GWs. Furthermore the term $R(T_{*})$ is the mean bubble separation evaluated at the percolation temperature, can be related to the inverse transition rate $\beta$, as
\begin{equation}
        R(T_{*}) =\frac{(8\pi)^\frac{1}{3}}{\beta} \max\left(c_{s},v_w\right).
\end{equation}
The term $\Tilde{\Omega}_{GW}\approx 10^{-2}$ stems from the numerical simulation as a residual prefactor, and the function $C(f/f_{p,0})$ describes the spectral shape
\begin{equation}
C(x) = x^{3}\left(\frac{7}{3 + 4 x^2}\right)^{\frac{7}{2}}\,,
\end{equation}
$f_{p,0}$ being the peak frequency redshifted until today
\begin{align}
f_{p,0} = & 26 \left(\frac{\beta}{v_w(8\pi)^{\frac{1}{3}}H(T_{*})}\right)\left(\frac{z_{p}}{10}\right) \times \nonumber \\
& \times \left(\frac{T_{*}}{100 \,\mathrm{GeV}}\right)\left(\frac{g(T_{*})}{100}\right)^{\frac{1}{6}} \mu \mathrm{Hz}\,.
\end{align}
The term $z_{p} \approx 10$ appears from the numerical simulation, and $T_{*}$ denote the temperature when the PT has completed and the GW's are being emitted at this temperature. The aforementioned set of equations are valid when the duration of the sound waves are less than a Hubble time.

\bibliography{Reference}
\end{document}